\documentstyle[12pt]{article}

\begin{document}
{\sf \begin{center} \noindent {\Large \bf Dynamos and anti-dynamos as thin magnetic flux ropes in Riemannian spaces}\\[3mm]

by \\[0.3cm]{\sl L.C. Garcia de Andrade}\\

\vspace{0.5cm} Departamento de F\'{\i}sica
Te\'orica -- IF -- Universidade do Estado do Rio de Janeiro-UERJ\\[-3mm]
Rua S\~ao Francisco Xavier, 524\\[-3mm]
Cep 20550-003, Maracan\~a, Rio de Janeiro, RJ, Brasil\\[-3mm]
Electronic mail address: garcia@dft.if.uerj.br\\[-3mm]
\vspace{2cm} {\bf Abstract}
\end{center}
\paragraph*{}
Two examples of magnetic anti-dynamos in magnetohydrodynamics (MHD)
are given. The first is a 3D metric conformally related to Arnold
cat fast dynamo metric:
${ds_{A}}^{2}=e^{-{\lambda}z}dp^{2}+e^{{\lambda}z}dq^{2}+dz^{2}$ is
shown to present a behaviour of non-dynamos where the magnetic field
exponentially decay in time.  The curvature decay as z-coordinates
increases without bounds. Some of the Riemann curvature components
such as $R_{pzpz}$ also undergoes dissipation while component
$R_{qzqz}$ increases without bounds. The remaining curvature
component $R_{pqpq}$ is constant on the torus surface. The other
anti-dynamo which may be useful in plasma astrophysics is the thin
magnetic flux rope or twisted magnetic thin flux tube which also
behaves as anti-dynamo since it also decays with time. This model is
based on the Riemannian metric of the magnetic twisted flux tube
where the axis possesses Frenet curvature and torsion. Since in this
last example the Frenet torsion of the axis of the rope is almost
zero, or the possible dynamo is almost planar it satisfies Zeldovich
theorem which states that planar dynamos do not exist. Changing in
topology of this result may result on a real dynamo as discussed.
\vspace{0.5cm} \noindent {\bf PACS numbers:}
\hfill\parbox[t]{13.5cm}{02.40.Hw-Riemannian geometries}

\newpage
\section{Introduction}
 Geometrical tools have been used with success \cite{1} in
 Einstein general relativity (GR) have been also used
 in other important areas of physics, such as plasma structures in tokamaks
 as been clear in the Mikhailovskii \cite{2} book to investigate the tearing
 and other sort of instabilities in confined plasmas \cite{2}, where the Riemann
 metric
 tensor plays a dynamical role interacting with the magnetic field through
 the magnetohydrodynamical equations (MHD). Recently
 Garcia de Andrade \cite{3,4} has also made use of Riemann metric to investigate
 magnetic flux tubes in superconducting plasmas.
 Thiffault and Boozer \cite{5} following the same reasoning applied the methods of Riemann
 geometry in the context of chaotic flows and fast dynamos. Yet more recently Thiffeault \cite{6} investigated the stretching and
 Riemannian curvature of material lines in chaotic flows as possible dynamos models. An interesting tutorial review of chaotic flows
 and kinematical dynamos has been presented earlier by Ott \cite{7}. Also Boozer \cite{8} has obtained a geomagnetic dynamo from conservation
 of magnetic helicity. This can also be shown here in the generalization to non-holonomic Frenet frame \cite{9}. In this paper we use the tools of Riemannian geometry,
 also user in GR, to obtain anti-dynamos in the conformal cat dynamo metric \cite{10}. We also use the Euler equations for
 incompressible flows in Arnold metric \cite{11}. Antidynamos or non-dynamos are also important in the respect that it is important
 to recognize when a topology or geometry of a magnetic field does force the field to decay exponentially for example. As we know
 planar dynamos does not exist and Anti-dynamos theorems are important in this respect. Thus in the present paper we also obtain
 antidynamos metrics which are conformally related to the fast dynamo metric discovered by Arnold.
 Levi-Civita connections \cite{12}  are found together Riemann curvature from the MAPLE X GR tensor
 package. The paper is organized as follows: In section II the
 the non-holonomic Frenet frame in MHD is briefly reviewed. In section III the flux rope thin antidynamo solution is
 presented.
 Curvature and connection are found for the Arnold metric in section IV. In section V the conformal antidynamo metric is considered. In
 section VI the dynamo solution is found from topological
 considerations. Conclusions are presented in section VII.
 \section{MHD scalar equations for kinematical dynamos in nonholonomic Frenet frame}
 Let us now start by considering the MHD field equations
\begin{equation}
{\nabla}.\vec{B}=0 \label{1}
\end{equation}
\begin{equation}
\frac{{\partial}}{{\partial}t}\vec{B}-{\nabla}{\times}[\vec{u}{\times}\vec{B}]-{\epsilon}{\nabla}^{2}\vec{B}=0
 \label{2}
\end{equation}
where $\vec{u}$ is a solenoidal field while ${\epsilon}$ is the
diffusion coefficient. Equation (\ref{2}) represents the induction
equation. The magnetic field $\vec{B}$ is chosen to lie along the
filament and is defined by the expression $\vec{B}=B({s,n})\vec{t}$
and $\vec{u}=u\vec{b}$ is the speed of the flow. The remaining
coordinate $n$ is orthogonal to the filament all along its
extension, and the arc length s measures distances along the the
filament itself. The vectors $\vec{t}$ and $\vec{n}$ along with
binormal vector $\vec{b}$ together form the Frenet frame which obeys
the Frenet-Serret equations
\begin{equation}
\vec{t}'=\kappa\vec{n} \label{3}
\end{equation}
\begin{equation}
\vec{n}'=-\kappa\vec{t}+ {\tau}\vec{b} \label{4}
\end{equation}
\begin{equation}
\vec{b}'=-{\tau}\vec{n} \label{5}
\end{equation}
the dash represents the ordinary derivation with respect to
coordinate s, and $\kappa(s,t)$ is the curvature of the curve where
$\kappa=R^{-1}$. Here ${\tau}$ represents the Frenet torsion. We
follow the assumption that the Frenet frame \cite{7} may depend on
other degrees of freedom such as that the gradient operator becomes
\begin{equation}
{\nabla}=\vec{t}\frac{\partial}{{\partial}s}+\vec{n}\frac{\partial}{{\partial}n}+\vec{b}\frac{\partial}{{\partial}b}
\label{6}
\end{equation}
 The other equations for the other legs of the Frenet frame are
\begin{equation}
\frac{\partial}{{\partial}n}\vec{t}={\theta}_{ns}\vec{n}+[{\Omega}_{b}+{\tau}]\vec{b}
\label{7}
\end{equation}
\begin{equation}
\frac{\partial}{{\partial}n}\vec{n}=-{\theta}_{ns}\vec{t}-
(div\vec{b})\vec{b} \label{8}
\end{equation}
\begin{equation}
\frac{\partial}{{\partial}n}\vec{b}=
-[{\Omega}_{b}+{\tau}]\vec{t}-(div{\vec{b}})\vec{n}\label{9}
\end{equation}
\begin{equation}
\frac{\partial}{{\partial}b}\vec{t}={\theta}_{bs}\vec{b}-[{\Omega}_{n}+{\tau}]\vec{n}
\label{10}
\end{equation}
\begin{equation}
\frac{\partial}{{\partial}b}\vec{n}=[{\Omega}_{n}+{\tau}]\vec{t}-
\kappa+(div\vec{n})\vec{b} \label{11}
\end{equation}
\begin{equation}
\frac{\partial}{{\partial}b}\vec{b}=
-{\theta}_{bs}\vec{t}-[\kappa+(div{\vec{n}})]\vec{n}\label{12}
\end{equation}
Another set of equations which we shall need here is the time
derivative of the Frenet frame given by
\begin{equation}
\dot{\vec{t}}=[{\kappa}'\vec{b}-{\kappa}{\tau}\vec{n}] \label{13}
\end{equation}
\begin{equation}
\dot{\vec{n}}={\kappa}\tau\vec{t} \label{14}
\end{equation}
\begin{equation}
\dot{\vec{b}}=-{\kappa}' \vec{t} \label{15}
\end{equation}
\section{Thin magnetic flux ropes as antidynamos in Riemannian spaces}
In this section we shall concerned with the presentation of a
solution of the dynamo equation investigated in previous section
which represents a antidynamo thin magnetic flux rope. Earlier
Yoshimura \cite{14} has investigated solar dynamos represented by
magnetic flux ropes which are actually another name for twisted
magnetic flux tubes. Let us now consider here the metric of magnetic
flux
\begin{equation}
ds^{2}=dr^{2}+r^{2}d{{\theta}_{R}}^{2}+{K^{2}}(s)ds^{2} \label{16}
\end{equation}
where the tube coordinates are $(r,{\theta}_{R},s)$ \cite{15} where
${\theta}(s)={\theta}_{R}-\int{{\tau}ds}$ where $\tau$ is the Frenet
torsion of the tube axis and $K(s)$ is given by
\begin{equation}
{K^{2}}(s)=[1-r{\kappa}(s)cos{\theta}(s)]^{2} \label{17}
\end{equation}
Computing the Riemannian Laplacian operator ${\nabla}^{2}$ in
curvilinear coordinates \cite{16} one obtains
\begin{equation}
{\nabla}^{2}=\frac{1}{\sqrt{g}}{\partial}_{i}[\sqrt{g}g^{ij}{\partial}_{j}]
\label{18}
\end{equation}
\begin{equation}
{\nabla}^{2}=
{{\partial}_{s}}^{2}+\frac{1}{r^{2}}{{\partial}_{{\theta}_{R}}}^{2}
-[\frac{{\theta}_{R}+{\tau}_{0}}{r}]{{\partial}}_{{\theta}_{R}}
\label{19}
\end{equation}
where ${\partial}_{j}:=\frac{{\partial}}{{\partial}x^{j}}$ and
$g:=det{g_{ij}}$ where $g_{ij}$ is the covariant component of the
Riemann metric of the flux rope. Applied to the above flux rope
coordinates this yields Note also that we have considered that the
flux rope magnetic field does not depend on the r and ${\theta}_{R}$
coordinates. This avoids the problem of changing poloidal components
into toroidal components of the field as in the dynamo solution of
Chatterjee, Choudhuri and Petroyav \cite{17} where his dynamo flux
tube considers radial components of the magnetic fields involved.
Thus the magnetic field here can be expressed as
\begin{equation}
\vec{B_{s}}=e^{pt}B_{0}\vec{t} \label{20}
\end{equation}
where from expression ${\nabla}.\vec{B}=0$ we note that $B_{0}$ is
constant, where in general $p=p(\epsilon)$. Thus substitution of
this expression in the equation for the dynamo we obtain
\begin{equation}
[{\kappa}_{0}{\tau}_{0}\vec{b}-{{\kappa}_{0}}^{2}\vec{t}]=p\vec{t}-{\kappa}_{0}[{\tau}_{0}+u_{0}]\vec{n}
\label{21}
\end{equation}
where the subscript zero indicates constant physical quantities, and
we consider the flow velocity $\vec{u}=u_{0}\vec{t}$ as constant in
modulus. Splitting this last expression component by component we
obtain the following scalar dynamo equations
\begin{equation}
{\tau}_{0}=-u_{0}\label{22}
\end{equation}
\begin{equation}
p=-{{\kappa}_{0}}^{2}{\epsilon} \label{23}
\end{equation}
\begin{equation}
{\kappa}_{0}{\tau}_{0}{\epsilon}=0 \label{24}
\end{equation}
These three equations altogether yield that very slow dynamos imply
that flux rope is almost planar and equation (\ref{23}) already
yields the following solution
\begin{equation}
\vec{B_{s}}=e^{-{{\kappa}_{0}}^{2}{\epsilon}t}B_{0}\vec{t}
\label{25}
\end{equation}
Since the resistivity ${\epsilon}{\ge}0$, this expression tells us
that the magnetic field cannot be sustained and decays in time which
shows clearly that this solution does not represent a dynamo. This
actually is in agreement with Zeldovich \cite{18} theorem since the
flux rope is almost planar.
\section{Riemann dynamos and dissipative manifolds and Euler flows}
 Arnold metric can be used to compute the Levi-Civita-Christoffel connection
\begin{equation}
 {{\Gamma}^{p}}_{pz}=-\frac{\lambda}{2}\label{26}
\end{equation}
\begin{equation}
 {{\Gamma}^{q}}_{qz}=\frac{\lambda}{2}\label{27}
\end{equation}
\begin{equation}
 {{\Gamma}^{z}}_{pp}=\frac{\lambda}{2}e^{-{\lambda}z}\label{28}
\end{equation}
\begin{equation}
 {{\Gamma}^{z}}_{qq}=-\frac{\lambda}{2}e^{-{\lambda}z}\label{29}
\end{equation}
from these connection components one obtains the Riemann tensor
components
\begin{equation}
 {R}_{pqpq}=-\frac{{\lambda}^{2}}{4}\label{30}
\end{equation}
Note that since this component is negative from the Jacobi equation
\cite{7} that the flow is unstable. The other components are
\begin{equation}
 {R_{pzpz}}=-\frac{{\lambda}^{2}}{2}e^{-{\lambda}z}\label{31}
\end{equation}
\begin{equation}
 R_{zqzq}=-\frac{{\lambda}^{2}}
 {2}e^{{\lambda}z}\label{32}
\end{equation}
one may immediatly notice that at large values of z the curvature
component $(zpzp)$ is bounded and vanishes,or undergoes a
dissipative effect, while component $(zqzq)$ of the curvature
increases without bounds, component $(pqpq)$ remains constant.
\section{Conformal anti-dynamo metric} Conformal metric techniques
have been widely used as a powerful tool obtain new solutions of the
Einstein's field equations of GR from known solutions. By analogy,
here we are using this method to yield new solutions of MHD
anti-dynamo solutions from the well-known fast dynamo Arnold
solution. We shall demonstrate that distinct physical features from
the Arnold solution maybe obtained. The conformal metric line
element can be defined as
\begin{equation}
ds^{2}={\lambda}^{-2z}{ds_{A}}^{2}={dx_{+}}^{2}+{\lambda}^{-4z}{dx_{-}}^{2}+{\lambda}^{-2z}dz^{2}\label{33}
\end{equation}
where we have used here the Childress and Gilbert \cite{5} notation
for the Arnold metric in ${\cal R}^{3}$ which reads now
\begin{equation}
{ds_{A}}^{2}={\lambda}^{2z}{dx_{+}}^{2}+{\lambda}^{-2z}{dx_{-}}^{2}+dz^{2}\label{34}
\end{equation}
where the coordinates are defined by
\begin{equation}
\vec{x}=x_{+}\vec{e_{+}}+x_{-}\vec{e_{-}}+z\vec{e}_{z}\label{35}
\end{equation}
where a right handed orthogonal set of vectors in the metric is
given by
\begin{equation}
\vec{f}_{+}=\vec{e}_{+} \label{36}
\end{equation}
\begin{equation}
\vec{f}_{-}={\lambda}^{2z}\vec{e}_{-} \label{37}
\end{equation}
\begin{equation}
\vec{f}_{z}={\lambda}^{z}\vec{e}_{z} \label{38}
\end{equation}
A component of a vector in this basis, such as the magnetic vector
$\vec{B}$ is
\begin{equation}
\vec{B}=B_{+}\vec{f}_{+}+B_{-}\vec{f}_{-}+B_{z}\vec{f}_{z}\label{39}
\end{equation}
The vector analysis formulas in this frame are
\begin{equation}
{\nabla}=[{\partial}_{+},{\lambda}^{2z}{\partial}_{-},{\lambda}^{z}{\partial}_{z}]\label{40}
\end{equation}
\begin{equation}
{\nabla}^{2}{\phi}=[{{\partial}_{+}}^{2}{\phi},{\lambda}^{4z}{{\partial}_{-}}^{2}{\phi},{\lambda}^{2z}{{\partial}_{z}}^{2}{\phi}]\label{41}
\end{equation}
The MHD dynamo equations are
\begin{equation}
{\nabla}.\vec{B}={{\partial}_{+}}B_{+}+{\lambda}^{2z}{{\partial}_{-}}B_{-}+{\lambda}^{z}{{\partial}_{z}}B_{z}=0\label{42}
\end{equation}
\begin{equation}
{\partial}_{t}\vec{B}+(\vec{u}.{\nabla})\vec{B}-(\vec{B}.{\nabla})\vec{u}={\epsilon}{\nabla}^{2}\vec{B}\label{43}
\end{equation}
where ${\epsilon}$ is the conductivity coefficient. Since here we
are working on the limit ${\epsilon}=0$ , which is enough to
understand the physical behavior of the fast dynamo, we do not need
to worry to expand the RHS of equation (\ref{43}), and it reduces to
\begin{equation}
(\vec{u}.{\nabla})\vec{B}={\partial}_{z}[B_{+}\vec{e}_{+}+B_{-}e^{2{\mu}z}\vec{e}_{-}+B_{z}e^{{\mu}z}\vec{e}_{z}]\label{44}
\end{equation}
where we have used that
$(\vec{B}.{\nabla})\vec{u}=B_{z}{\mu}e^{{\mu}z}\vec{e}_{z}$ and that
${\mu}=log{\lambda}$. This is one of the main differences between
Arnold metric and ours since in his fast dynamo, this relation
vanishes since in Arnold metric $\vec{u}=\vec{e}_{z}$ where
$\vec{e}_{z}$ is part of a constant basis. Separating the equation
in terms of the coefficients of $\vec{e}_{+}$, $\vec{e}_{-}$ and
$\vec{e}_{z}$ respectively one obtains the following scalar
equations
\begin{equation}
{\partial}_{z}B_{+}+{\partial}_{t}B_{+}=0\label{45}
\end{equation}
\begin{equation}
{\partial}_{t}B_{-}+{\partial}_{t}B_+2{\mu}B_{-}=0\label{46}
\end{equation}
\begin{equation}
{\partial}_{t}B_{z}+{\partial}_{z}B_=0\label{47}
\end{equation}
Solutions of these equations allows us to write down an expression
for the magnetic vector field $\vec{B}$ as
\begin{equation}
\vec{B}=[{B^{0}}_{z},{\lambda}^{-(t+z)}{B^{0}}_{-},{B^{0}}_{z}](t-z,y,x+y)\label{48}
\end{equation}
From this expression we can infer that the field is carried in the
flow, stretched in the $\vec{f}_{z}$ direction and compressed in the
$\vec{f}_{-}$ direction, while in Arnold's cat fast dynamo is also
compressed along the $\vec{f}_{-}$ direction but is stretched along
$\vec{f}_{+}$ direction while here this direction is not affected.
But the main point of this solution is the fact that the solution
represents an anti-dynamo since as one can see from expression
(\ref{48}) the magnetic field fastly decays exponentially in time as
$e^{{\mu}(t+z)}$. Let us now compute the Riemann tensor components
of the new conformal metric to check for the stability of the
non-dynamo flow. To easily compute this curvature components we
shall make use of Elie Cartan \cite{13} calculus of differential
forms, which allows us to express the conformal metric as
\begin{equation}
ds^{2}={dp}^{2}+e^{4{\lambda}z}{dq}^{2}+e^{{\lambda}z}dz^{2}\label{49}
\end{equation}
or in terms of the frame basis form ${\omega}^{i}$ is
\begin{equation}
ds^{2}=({{\omega}^{p}})^{2}+({{\omega}^{q}})^{2}+({{\omega}_{z}})^{2}\label{50}
\end{equation}
where we are back to Arnold's notation for convenience. The basis
form are write as
\begin{equation}
{\omega}^{p}=dp \label{51}
\end{equation}
\begin{equation}
{\omega}^{q}=e^{{\lambda}z}dq \label{52}
\end{equation}
and
\begin{equation}
{\omega}^{z}=e^{{\frac{\lambda}{2}}z}dq \label{53}
\end{equation}
By applying the exterior differentiation in this basis form one
obtains
\begin{equation}
d{\omega}^{p}=0 \label{54}
\end{equation}
\begin{equation}
d{\omega}^{z}=0 \label{55}
\end{equation}
and
\begin{equation}
d{\omega}^{q}={\lambda}e^{-{\frac{\lambda}{2}}z}{\omega}^{z}{\wedge}{\omega}^{q}
\label{56}
\end{equation}
Substitution of these expressions into the first Cartan structure
equations one obtains
\begin{equation}
T^{p}=0={{\omega}^{p}}_{q}{\wedge}{\omega}^{q}+
{{\omega}^{p}}_{z}{\wedge}{\omega}^{z}\label{57}
\end{equation}
\begin{equation}
T^{q}=0={\lambda}e^{-{\frac{\lambda}{2}}z}{\omega}^{z}{\wedge}{\omega}^{q}+{{\omega}^{q}}_{p}{\wedge}{\omega}^{p}+{{\omega}^{q}}_{z}{\wedge}{\omega}^{z}
\label{58}
\end{equation}
and
\begin{equation}
T^{z}=0={{\omega}^{z}}_{p}{\wedge}{\omega}^{p}+{{\omega}^{z}}_{q}{\wedge}{\omega}^{q}
\label{59}
\end{equation}
where $T^{i}$ are the Cartan torsion 2-form which vanishes
identically on a Riemannian manifold. From these expressions one is
able to compute the connection forms which yields
\begin{equation}
{{\omega}^{p}}_{q}=-{\alpha}{\omega}^{p}\label{60}
\end{equation}
\begin{equation}
{{\omega}^{q}}_{z}={\lambda}e^{-{\frac{\lambda}{2}}z}{\omega}^{q}
\label{62}
\end{equation}
and
\begin{equation}
{{\omega}^{z}}_{p}={\beta}{\omega}^{p} \label{63}
\end{equation}
where ${\alpha}$ and ${\beta}$ are constants. Substitution of these
connection form into the second Cartan equation
\begin{equation}
{R^{i}}_{j}={R^{i}}_{jkl}{\omega}^{k}{\wedge}{\omega}^{l}=d{{\omega}^{i}}_{j}+{{\omega}^{i}}_{l}{\wedge}{{\omega}^{l}}_{j}
\label{64}
\end{equation}
where ${R^{i}}_{j}$ is the Riemann curvature 2-form. After some
algebra we obtain the following components of Riemann curvature for
the conformal antidynamo
\begin{equation}
{R^{p}}_{qpq}= {\lambda}e^{-{\frac{\lambda}{2}}z}\label{65}
\end{equation}
\begin{equation}
{R^{q}}_{zqz}= \frac{1}{2}{\lambda}^{2}e^{-{\lambda}z}\label{66}
\end{equation}
and finally
\begin{equation}
{R^{p}}_{zpq}= -{\alpha}{\lambda}e^{-{\frac{\lambda}{2}}z}\label{67}
\end{equation}
We note that only component to which we can say is positive is
${R^{p}}_{zqz}$ which turns the flow stable in this q-z surface.
This component also dissipates away when $z$ increases without
bounds, the same happens with the other curvature components
\cite{13}. \section{Dynamos by topology change}
 A long and
straithforward computation ,specially due to the computation of
${\nabla}^{2}A$. and substituting these equations for the dynamics
of the Frenet frame leads to the scalar MHD expressions
\begin{equation}
{\partial}_{t}A=-{\partial}_{s}{\phi}+[{{\partial}^{2}}_{n}A-A({{\theta}_{ns}}^{2}-{{\kappa}_{0}}^{2})]
\label{68}
\end{equation}
\begin{equation}
-{\kappa}{\tau}A=-uB+{\epsilon}[2{\partial}_{n}A+({\Omega}_{s}+{\tau}){\theta}_{ns}A]
\label{69}
\end{equation}
\begin{equation}
-{\theta}_{bs}A={\epsilon}[2{\partial}_{n}A{\Omega}_{s}+{\Omega}^{2}A]
\label{70}
\end{equation}
where ${\kappa}_{0}$ is the Frenet curvature of the streamlines.
These equations have already been simplified by using the relations
\begin{equation}
{\nabla}{\times}\vec{A}=\vec{B}\label{71}
\end{equation}
which yields the following differential scalar equations
\begin{equation}
{B}=-A[{\Omega}_{b}+\tau] \label{72}
\end{equation}
\begin{equation}
{\partial}_{n}A+{\kappa}A=0 \label{73}
\end{equation}
\begin{equation}
A({\Omega}_{n}+{\tau})=0 \label{74}
\end{equation}
Where the ${\Omega}'s$ represent the abnormalities of the
streamlines of the flow. When the ${\Omega}_{s}$ vanishes we note
the geodesic streamlines are obtained. As we shall see below here we
are not consider geodesic flows dynamos. By considering planar flows
where torsion vanishes and the gauge condition
\begin{equation}
{\nabla}.\vec{A}+\frac{\partial}{{\partial}t}{\phi}=0 \label{75}
\end{equation}
This equation can be expressed as
\begin{equation}
{\partial}_{s}{A}+[{\theta}_{ns}+{\theta}_{bs}]A=0 \label{76}
\end{equation}
Now by considering that A does not depend on the coordinate s this
expression reduces to
\begin{equation}
[{\theta}_{ns}+{\theta}_{bs}]A=0 \label{77}
\end{equation}
which reduces to ${\theta}_{ns}=-{\theta}_{bs}$. By making use of
this expression and the assumption that ${\phi}=0$ one simplifies
the MHD scalar equations to
\begin{equation}
{\partial}_{t}A=[{{\partial}^{2}}_{n}A-A({{\theta}_{ns}}^{2}-{{\kappa}_{0}}^{2})]
\label{78}
\end{equation}
\begin{equation}
uA{\Omega}_{b}+{\epsilon}[2{\kappa}_{0}+{\Omega}_{s}]A=0 \label{79}
\end{equation}
\begin{equation}
{\theta}_{ns}A={\epsilon}[-2{\kappa}_{0}{\Omega}_{s}+{{\Omega}_{s}}
^{2}]{\theta}_{ns}A \label{79}
\end{equation}
Simple solution of these two last equations reads
\begin{equation}
u=-\frac{{\Omega}_{s}{\epsilon}^{2}{{\kappa}_{0}}^{2}}{{\Omega}_{b}}
\label{80}
\end{equation}
and
\begin{equation}
{\theta}_{ns}=[b^{2}-2{\kappa}_{0}b] \label{81}
\end{equation}
where $b:={\Omega}_{s}$. FRom expression
\begin{equation}
{\partial}_{t}A-{{\theta}_{ns}}^{2}A=0 \label{82}
\end{equation}
which yields the solution
\begin{equation}
A=A_{0}(n)e^{[{\epsilon}{{\theta}_{ns}}^{2}]t} \label{83}
\end{equation}
To simplify the analysis of Arnold's theorem \cite{8} in the next
section we consider the geodesic flow assumption which simplifies
this solution to $A=A_{0}(n)$ which by solving the equation
(\ref{82}) yields
\begin{equation}
A={{A_{0}}^{*}}e^{-[{\kappa}_{0}]n} \label{84}
\end{equation}
and finally we note that the magnetic field of streamlines becomes
\begin{equation}
B=-{\Omega}_{b}{{A_{0}}^{*}}e^{-[{\kappa}_{0}]n} \label{85}
\end{equation}
We note from this expression that if the signs of Frenet curvature
and coordinate n coincides the magnetic field decays in space and a
kinematical dynamo is not obtained. However if the signs do not
agree such as ${\kappa}_{0}>0$ (positive curvature of the
streamlines) and $n<0$ the magnetic field increases with the
distance from the streamlines and a kinematical dynamo is obtained.
\section{Conclusions}
 In conclusion, we have used a well-known technique to find solutions of Einstein's field equations of gravity
 namely the conformal related spacetime metrics to find a new anti-dynamo solution in MHD nonplanar flows.
 The stability of the flow is also analysed by using other tools from
 GR, namely that of Killing symmetries. Examination of the  Riemann curvature components enable one to
 analyse the stretch and compression of the dynamo flow. Other interesting antidynamo metric
 equations in four-dimensional spacetime \cite{1}. \section*{Acknowledgements}
 Thanks are due to CNPq and UERJ for financial supports. of a thin
 almost planar flux rope is also shown. Possibility of obtaining
 dynamos by changing topology in the nonholonomy Frenet frame is
 also discussed. It is shown that in this case dynamo solution is a
 realistic possibility.

\end{document}